# Why the traditional concept of local hardness does not work

Tamás Gál*

Quantum Theory Project, Department of Physics, University of Florida,

Gainesville, Florida 32611, USA

**Abstract:** Finding a proper local measure of chemical hardness has been a long-standing aim of density functional theory. The traditional approach to defining a local hardness index, by the derivative of the chemical potential $\mu$ with respect to the electron density $n(\vec{r})$ subject to the constraint of a fixed external potential $v(\vec{r})$, has raised several questions, and its chemical applicability has proved to be limited. Here, we point out that the only actual possibility to obtain a local hardness measure in the traditional approach emerges if the external potential constraint is dropped; consequently, utilizing the ambiguity of a restricted chemical potential derivative is not an option to gain alternative definitions of local hardness. At the same time, however, the arising local hardness concept turns out to be fatally undermined by its inherent connection with the asymptotic value of the second derivative of the universal density functional. The only other local hardness concept one may deduce from the traditional definition $\delta\mu[n]/\delta n(\vec{r})\big|_{v(\vec{r})}$ is the one that gives a constant value, the global hardness itself, throughout an electron system in its ground state. Consequently, the traditional approach is *in principle* incapable of delivering a local hardness indicator. The parallel case of defining a local version of the chemical potential itself is also outlined, arriving at a similar conclusion. Namely, the only local chemical potential concept that can be gained from a definition $\delta E[n]/\delta n(\vec{r})\big|_{v(\vec{r})}$ is the one that gives a constant, $\mu$ itself, for electron systems in their ground state.

*E-mail: gal@qtp.ufl.edu

# I. Introduction

Chemical reactivity indices [1,2], defined within the framework of density functional theory (DFT) [1], have found successful application in the study of chemical phenomena. The three most well known reactivity descriptors, the electronegativity [3-6], or in the language of DFT, minus the chemical potential [6], the chemical hardness, and its inverse, the softness [7-10], are basic constituents of essential principles governing chemical reactions – the electronegativity equalization principle [6,11], the hard/soft acid/base principle [7-10,12-17], and the maximum hardness principle [18-23]. An important aim of chemical reactivity theory [2] is to establish local versions of the global indices, on the basis of which predictions can be made regarding the molecular sites a given reaction happens at.

Defining a local softness can be done in a natural way [24], by replacing the electron number $N$ with the electron density $n(\vec{r})$ in the definition of softness as the derivative of $N$ with respect to the chemical potential $\mu$. On the other hand, defining a local counterpart [25,26] of hardness, the multiplicative inverse of softness, has met essential difficulties [27-33], which undermine the applicability of the local hardness concept. This may not seem to be a substantial problem, as the concepts of hardness and softness are simple complementers; however, very recently, even the definition of local softness $s(\vec{r})$ has been found to fail to properly signify the soft sites in the case of hard systems [34,35], and even before, the interpretation of small $s(\vec{r})$ values as indicators of locally hard sites, preferred in hard-hard interactions [36], had been put into question [37-39]. Therefore, the question of a possible existence of a proper local hardness indicator has a renewed significance. In this study, we will re-examine the idea of defining a local hardness concept via differentiation of the chemical potential with respect to the density subject to the constraint of a fixed external potential $v(\vec{r})$, in the view of the questions as to (i) why this traditional way of defining a local hardness concept could not yield a (generally) correct local indicator of chemical hardness, and (ii) whether there is any possibility to gain such a local index from this approach. We will find that the only possibility to obtain a proper local hardness measure in the traditional way emerges if the external potential constraint on the differentiation is dropped. The evaluation of the arising local hardness concept, however, will be shown to be fatally undermined by the necessary involvement of the asymptotic fixation of the external potential. At the same time, we will show that the constant local hardness of Ghosh [26] emerges as the unique constrained derivative corresponding to the fixed-$v(\vec{r})$ constraint – but this local hardness concept cannot be a local reactivity measure because of its constancy. Our conclusion will be that the traditional approach to defining a local hardness index is not capable of delivering a proper local hardness measure; therefore, an essentially new approach to this problem needs to be found (like that proposed in [33], which originates a local hardness index via a local chemical potential – a chemical potential density). We will also consider the analogous case of defining a local counterpart of the chemical potential itself, having relevance (i) regarding the definition of a local electronegativity index and (ii) serving as a potential basis for an alternative local hardness definition. The results will be similar to the local hardness case – in particular, the idea of defining a local chemical potential by the derivative of the ground-state energy with respect to the density subject to the constraint of a fixed external potential yields the constant local chemical potential concept of Parr et al. [6] as the only feasible definition.

## II. The traditional concept of local hardness

The chemical concept of hardness has been quantified by Parr and Pearson [8] as

$$\eta = \left(\frac{\partial \mu}{\partial N}\right)_{v(\vec{r})} . \qquad (1)$$

In contrast with its inverse chemical quantity, the softness

$$S = \left(\frac{\partial N}{\partial \mu}\right)_{v(\vec{r})} , \qquad (2)$$

defining a local counterpart for hardness has met essential difficulties, due to the fact that there is no such obvious way to do this as in the case of Eq.(2). For Eq.(2), a corresponding local quantity can be readily introduced [24]:

$$s(\vec{r}) = \left(\frac{\partial n(\vec{r})}{\partial \mu}\right)_{v(\vec{r})} , \qquad (3)$$

which has been termed local softness. This has a direct connection to the Fukui function [40]

$$f(\vec{r}) = \left(\frac{\partial n(\vec{r})}{\partial N}\right)_{v(\vec{r})} , \qquad (4)$$

a well-established chemical reactivity index: Applying the chain rule of differentiation, one obtains

$$s(\vec{r}) = \left(\frac{\partial n(\vec{r})}{\partial N}\right)_{v(\vec{r})} \left(\frac{\partial N}{\partial \mu}\right)_{v(\vec{r})} = f(\vec{r}) S . \qquad (5)$$

$s(\vec{r})$ integrates to $S$ (just as the Fukui function integrates to 1), and it is natural to interpret it as a pointwise, i.e. local, softness [24].

A local hardness concept was first introduced by Berkowitz et al. [25], who defined the local hardness as

$$\eta(\vec{r}) = \left(\frac{\delta\mu}{\delta n(\vec{r})}\right)_{v(\vec{r})} . \qquad (6)$$

This local index is then not a local quantity in the sense the local softness is, since it does not integrate to the hardness; consequently, its integral over a given region in the molecule won't give a regional global hardness. In fact, $\eta(\vec{r})$ times the Fukui function is what gives $\eta$ by integration over the whole space,

$$\int \eta(\vec{r}) f(\vec{r}) d\vec{r} = \eta , \qquad (7)$$

which emerges via an application of the chain rule, as can be seen from the definitions Eqs.(4) and (6).

The biggest difficulty with the local hardness defined by Eq.(6) has been that it is not clear how to understand the fixed external potential [$v(\vec{r})$] condition on the differentiation in Eq.(6). If we consider that the hardness is defined by Eq.(1) as the partial derivative of the chemical potential $\mu[N,v]$ (a function(al) of the electron number and the external potential) with respect to $N$, Eq.(6) suggests that $v(\vec{r})$ *as one of the variables* in $\mu[N,v]$ should be fixed when differentiating with respect to the electron density $n(\vec{r})$. However, this yields

$$\eta(\vec{r}) = \left(\frac{\partial \mu[N,v]}{\partial N}\right)_{v(\vec{r})} \frac{\delta N}{\delta n(\vec{r})} = \eta , \qquad (8)$$

i.e., the local hardness equals the global hardness at every point in space. If one utilizes the DFT Euler-Lagrange equation

$$\frac{\delta F[n]}{\delta n(\vec{r})} + v(\vec{r}) = \mu , \qquad (9)$$

emerging from the minimization principle for the ground-state energy density functional

$$E_v[n] = F[n] + \int n(\vec{r}) v(\vec{r}) d\vec{r} \qquad (10)$$

for the determination of the ground-state density corresponding to a given $v(\vec{r})$, $\mu[N,v]$ can be given as

$$\mu[N,v] = \frac{\delta F}{\delta n(\vec{r})}[n[N,v]] + v(\vec{r}) . \qquad (11)$$

Differentiating this expression with respect to $N$ yields

$$\eta = \left(\frac{\partial \mu[N,v]}{\partial N}\right)_v = \int \frac{\delta^2 F}{\delta n(\vec{r})\delta n(\vec{r}')} \left(\frac{\partial n(\vec{r}')}{\partial N}\right)_v d\vec{r}' = \int \frac{\delta^2 F}{\delta n(\vec{r})\delta n(\vec{r}')} f(\vec{r}') d\vec{r}' \ . \tag{12}$$

On the basis of this, then, it is natural to identify the local hardness yielding Eq.(8) with

$$\eta(\vec{r}) = \int \frac{\delta^2 F}{\delta n(\vec{r})\delta n(\vec{r}')} f(\vec{r}') d\vec{r}' \ . \tag{13}$$

Eq.(13) was proposed by Ghosh [26], and was discovered to be a constant giving the global hardness everywhere (for the ground-state density) by Harbola et al. [27]. Eq.(13) thus cannot be a local counterpart of hardness on the basis of which one could differentiate between molecular sites; however, it still is a useful conceptual and practical tool since a local hardness equalization principle can be based on it [41-44], which says that $\eta(\vec{r})$ of Eq.(13) should be constant for the whole molecule for the real ground-state density – but only for that density. This principle is closely related with the long-known chemical potential (or electronegativity) equalization principle [6,43,45].

To gain other definition for the local hardness than the one yielding the global hardness in every point of space, one may consider the fixed-$v(\vec{r})$ constraint in Eq.(6) as a constraint on the differentiation with respect to the density,

$$\eta(\vec{r}) = \left.\frac{\delta \mu[N[n],v[n]]}{\delta n(\vec{r})}\right|_{v(\vec{r})} , \tag{14}$$

instead of a simple fixation of the variable $v(\vec{r})$ of $\mu[N,v]$. That is, the density domain over which the differentiation is carried out is restricted to the domain of densities that yield the given $v(\vec{r})$, through the first Hohenberg-Kohn theorem [1], which constitutes a unique $n(\vec{r}) \to v(\vec{r})$ mapping, i.e. a $v(\vec{r})[n]$ functional. The result will be an ambiguous restricted derivative (see Sec.II of [46]), similarly to the case of derivatives restricted to a domain of densities with a given normalization $N$, which derivatives are determined only up to an arbitrary additive constant [1,47].

Harbola et al. [27], to characterize the ambiguity of the local hardness concept of Eq.(6), first recognized by Ghosh [26], have given the explicit form

$$\eta(\vec{r}) = \int \frac{\delta^2 F}{\delta n(\vec{r})\delta n(\vec{r}')} u(\vec{r}') d\vec{r}' \tag{15}$$

for the possible local hardness candidates, where $u(\vec{r})$ is an arbitrary function that integrates to 1. The second derivative of $F[n]$, appearing in Eq.(15), is called the hardness kernel [48], which also serves as a basis for a minimization theorem determining the fukui function [49].

The choice $u(\vec{r}) = f(\vec{r})$ gives back Eq.(13), while another natural choice is $u(\vec{r}) = n(\vec{r})/N$, which yields the original local hardness formula of Berkowitz et al. [25],

$$\eta(\vec{r}) = \frac{1}{N} \int \frac{\delta^2 F}{\delta n(\vec{r}) \delta n(\vec{r}')} n(\vec{r}') d\vec{r}' , \qquad (16)$$

who deduced it as a consequence of Eq.(6).

Besides the above two definitions for $\eta(\vec{r})$, another one, termed the unconstrained local hardness, has been proposed by Ayers and Parr [21,31]:

$$\eta(\vec{r}) = \frac{\delta \mu[N[n], v[n]]}{\delta n(\vec{r})} , \qquad (17)$$

where the fixed-$v(\vec{r})$ constraint on the differentiation with respect to $n(\vec{r})$ is simply dropped. A substantial difficulty with this definition as regards practical use [31] is the explicit appearance of the derivative of $v(\vec{r}')$ with respect to $n(\vec{r})$, as can be seen by

$$\eta(\vec{r}) = \frac{\partial \mu}{\partial N} + \int \frac{\delta \mu[N,v]}{\delta v(\vec{r}')} \frac{\delta v(\vec{r}')}{\delta n(\vec{r})} d\vec{r}' = \eta + \int f(\vec{r}') \frac{\delta v(\vec{r}')}{\delta n(\vec{r})} d\vec{r}' , \qquad (18)$$

where the well-known fact

$$\frac{\delta E[N,v]}{\delta v(\vec{r})} = n(\vec{r}) \qquad (19)$$

and Eqs.(1) and (4) have been utilized. Note that Eq.(17), too, is embraced by Eq.(14), since for a restricted derivative, a trivial choice is the unrestricted derivative itself (if exists), being valid over the whole functional domain, hence over the restricted domain too.

### III. Excluding the ambiguity of the local hardness concept of Eq.(14)

A proper local hardness is expected to yield proper regional hardness values, on the basis of which one can predict the molecular region (or site) a reaction with another species happens at. The only known and plausible way of obtaining regional hardnesses from an $\eta(\vec{r})$ defined by Eq.(14) is

$$\eta_\Omega = \int_\Omega \eta(\vec{r}) \left( \frac{\partial n(\vec{r})}{\partial N} \right)_v d\vec{r} = \int_\Omega \eta(\vec{r}) f(\vec{r}) d\vec{r} ; \qquad (20)$$

i.e., the integral in Eq.(7) is carried out over the considered region $\Omega$ of space instead of the whole space. Eq.(20) has been applied in practical calculations to characterize the hardness of atomic regions or functional groups in molecules [50], and as a special case (in the form of Eq.(7)), to evaluate the global hardness itself [51]. Eq.(20) represents an "extensive" hardness concept: The total hardness of a molecule can be obtained as a sum of its regional hardnesses corresponding to a given (arbitrary) division of the molecule into regions. That is, roughly saying, a molecule that contains regions having high values of hardnesses in a majority will have a high global hardness, while a molecule that contains mainly soft regions, with low $\eta(\Omega)$, will have a low global hardness. Of course, in a strict sense, the hardness won't be an extensive property, since for the determination of the hardness of a given region, the whole of the electronic system needs to be involved (a change in the electron number induces a change in the electron density distribution as a whole) – however, we cannot expect more in quantum mechanics, since there is no sense in asking how much a given property of a segment of a system changes due to the addition of a fraction d$N$ of electrons to, and only to, that segment.

A problem with this local hardness/regional hardness scheme is that if $\eta$ is extensive, with regional hardnesses given by Eq.(20), the quantity $\eta(\vec{r})f(\vec{r})$ should be considered to be the local hardness instead of $\eta(\vec{r})$ (of Eq.(14)). A local quantity $\rho(\vec{r})$ corresponding to a given extensive global quantity $\Theta$ emerges as $\rho(\vec{r}) = \lim_{\Delta V \to \vec{r}} \frac{\Delta \Theta}{\Delta V}$, implying $\Theta(\Omega) = \int_\Omega \rho(\vec{r}) d\vec{r}$. At the same time, however, it seems plausible to take $\eta(\vec{r})$ of Eq.(14) as the local hardness since it characterizes the change of the chemical potential induced by a small (infinitesimal) change of the electron density $n(\vec{r})$ at a given point of space in a given external potential – this seems to be a proper local counterpart of the hardness, given by Eq.(1). Although this view is intuitively appealing, one should be careful with such an approach, because then we may argue that a change (even if infinitesimal) of the density at a single $\vec{r}$ will yield a discontinuous density, so why should one bother himself with chemical potential changes corresponding to unphysical density changes? This point is just to show the dangerous side of intuitive arguing – but there is a more physical/real argument against an $\eta(\vec{r})$ directly defined by Eq.(14). If we add a small fraction d$N$ of number of electrons to a molecule, it will be distributed over the whole molecule, no matter "where we added" that d$N$ of electrons. Consequently, only a change of $\mu$ that is induced by a density change that is caused by a d$N$ makes sense directly. $\frac{\delta \mu}{\delta n(\vec{r})}$ is only an intermediate quantity that delivers the infinitesimal change in $\mu$ due to an infinitesimal change of $N$ or some other quantity determining the given electron system and hence $n(\vec{r})$ – as

$$\left(\frac{\partial \mu}{\partial N}\right)_v = \int \left.\frac{\delta \mu}{\delta n(\vec{r})}\right|_v \left(\frac{\partial n(\vec{r})}{\partial N}\right)_v d\vec{r} \ , \tag{21}$$

or

$$\left(\frac{\delta\mu}{\delta v(\vec{r})}\right)_N = \int \left.\frac{\delta\mu}{\delta n(\vec{r}')}\right|_N \left(\frac{\delta n(\vec{r}')}{\delta v(\vec{r})}\right)_N d\vec{r}' , \quad (22)$$

e.g. (Provided it exists, an unrestricted derivative of $\mu$ with respect to $n(\vec{r})$, $\frac{\delta\mu}{\delta n(\vec{r})}$, may be used in both of the above equations in the place of the restricted derivatives.) Thus, it may be more appropriate to term Eq.(14) e.g. as "local hardness factor", instead of "local hardness", which indicates its role in delivering the actual local hardness $\eta(\vec{r})f(\vec{r})$ and regional hardnesses. Of course, this is just a matter of terminology (and why should we change a name nearly 30 years old?); however, the relevant point here is that one should not expect $\eta(\vec{r})$ of Eq.(14) itself to be a measure of local hardness – at least if one expects Eq.(20) to deliver regional hardness measures. The question of considering $\eta(\vec{r})f(\vec{r})$ a local hardness measure instead of $\eta(\vec{r})$ was first raised by Langenaeker et al. [52] (to get a proper complementer quantity of local softness $s(\vec{r})$), who called $\eta(\vec{r})f(\vec{r})$ "hardness density". The latter term, of course, is an appropriate name for $\eta(\vec{r})f(\vec{r})$ as this integrates to the hardness, and even more appropriate for an extensive quantity; however, if $\eta(\vec{r})f(\vec{r})$ proved to be a proper hardness density distribution indeed (with larger values in harder regions), it should also be termed "local hardness", since it would then be a local measure of hardness. But if (some choice of) $\eta(\vec{r})$ of Eq.(14) *itself* turned out to be a proper local hardness measure, it is $\eta(\vec{r})$ what should be termed "local hardness" (and terming $\eta(\vec{r})f(\vec{r})$ "hardness density", just because it integrates to the hardness, would become strongly questionable). $\eta(\vec{r})$ and $\eta(\vec{r})f(\vec{r})$ simultaneously cannot be a correct measure of local hardness. The question as to whether $\eta(\vec{r})$ of Eq.(14) itself may be a local hardness indicator will be examined in Sec. V. Note that the local hardness does not *have to* be a property density [53] – but if Eq.(20) is to deliver regional hardnesses, then it does have to be, and then it cannot be $\eta(\vec{r})$ of Eq.(14) itself.

Now, the question is as to whether any choice of Eq.(14), i.e. any way of fixing the external potential while differentiating with respect to the density, is indeed an allowed choice to obtain a local quantity $\eta(\vec{r})f(\vec{r})$ that correctly delivers regional hardnesses. The answer is negative; the only possible concrete choice of Eq.(14) is the unconstrained local hardness (factor) of Ayers and Parr, as we will see. Consider Eqs.(21) and (22) with the integrals taken only over a given region of space. We are interested (directly) only in the case of Eq.(21), but by the example of Eq.(22), more insight may be gained; therefore, it is worth considering it, too, in parallel with Eq.(21). Thus, we have, on one hand, Eq.(20), and on the other hand,

$$f_\Omega(\vec{r}) = \int_\Omega \left.\frac{\delta\mu}{\delta n(\vec{r}')}\right|_N \left(\frac{\delta n(\vec{r}')}{\delta v(\vec{r})}\right)_N d\vec{r}' , \qquad (23)$$

which is a "regional Fukui function", as the left-hand side of Eq.(22) is just the Fukui function,

$$\left(\frac{\delta\mu}{\delta v(\vec{r})}\right)_N = f(\vec{r}) , \qquad (24)$$

due to Eq.(19). What do these regional integrals tell us? They can be viewed as entities that give the contributions, to the infinitesimal change of $\mu$, that come from the change of the density over the given region due to an increment of $N$ and $v(\vec{r})$, respectively. To ease understanding, compare this with the finite-dimensional example of a function $g(x(t), y(t))$ (with a derivative $\dot{g} = \frac{\partial g}{\partial x}\frac{dx}{dt} + \frac{\partial g}{\partial y}\frac{dy}{dt}$, with respect to $t$), for which a "regional integral", or partial sum, means $\dot{g}_x = \frac{\partial g}{\partial x}\frac{dx}{dt}$ – that is, the part of $\dot{g}$ that is due to the $x$ part of the full change of $g(x(t), y(t))$ with respect to $t$. Thus, an infinitesimal change of $N$, or $v(\vec{r})$, induces a density change $\delta n(\vec{r})$, then the regional integral Eq.(20), or Eq.(23), tells us how much the part of $\delta n(\vec{r})$ that falls on the given domain $\Omega$ contributes to the whole change $\partial\mu$ of $\mu$ due to $\delta n(\vec{r})$, via $(\partial\mu)_\Omega = \int_\Omega \frac{\delta\mu}{\delta n(\vec{r})} \delta n(\vec{r}) d\vec{r}$. This indeed is a plausible way to obtain a regional hardness measure – but *only* if the unrestricted (i.e., full) derivative of $\mu$ is applied, as will be pointed out below. Note that Eq.(23) gives a natural decomposition of the Fukui function $f(\vec{r})$, $\sum_i f_{\Omega_i}(\vec{r}) = f(\vec{r})$. It gives how much contribution to $f(\vec{r})$, at any given $\vec{r}$, can be attributed to a given region $\Omega$ of the molecule (not necessarily including $\vec{r}$ !).

To understand why the full derivative $\frac{\delta\mu}{\delta n(\vec{r})}$ is the only possible choice in Eqs.(20) and (23) to yield proper regional measures, it is important to see where the ambiguity of restricted derivatives emerges from. The derivative of a functional $A[\rho]$, as used in physics, is defined by

$$\int \frac{\delta A[\rho]}{\delta\rho(x')} \Delta\rho(x') dx' = D(A)[\rho; \Delta\rho] , \qquad (25a)$$

which has to hold for any $\Delta\rho(x)$, and where $D(A)[\rho;\Delta\rho]$ denotes the Fréchet, or Gâteaux, differential of $A[\rho]$ for $\Delta\rho(x)$; see [46] for details. Eq.(25) may be written less rigorously as

$$\int \frac{\delta A[\rho]}{\delta\rho(x')} \delta\rho(x')\,dx' = A[\rho+\delta\rho] - A[\rho] \,, \tag{25b}$$

where $\delta\rho(x)$ denotes a first-order, i.e. "infinitesimal", increment of $\rho(x)$. Now, if we restrict the functional domain by the requirement that the $\rho(x)$'s of the domain have to satisfy some constraint $C[\rho] = C$ (i.e., we are not expecting the functional derivative to be valid over the whole domain of $\rho(x)$'s), this implies that more than one function $\frac{\delta A[\rho]}{\delta\rho(x)}$ is capable of delivering $D(A)[\rho;\Delta\rho]$ for any $\Delta\rho(x)$ that is in accordance with the constraint, $\Delta_C \rho(x)$. Namely, if $\frac{\delta A[\rho]}{\delta\rho(x)}$ fulfils Eq.(25), any other $\frac{\delta A[\rho]}{\delta\rho(x)} + \lambda \frac{\delta C[\rho]}{\delta\rho(x)}$ will fulfill it, too, over the given restricted domain, since

$$\int \frac{\delta C[\rho]}{\delta\rho(x')} \delta_C \rho(x')\,dx' = 0 \,, \tag{26}$$

emerging from $C[\rho+\delta_C\rho] - C[\rho] = 0$. Denoting a restricted derivative by $\left.\frac{\delta A[\rho]}{\delta\rho(x)}\right|_C$, while reserving the notation $\frac{\delta A[\rho]}{\delta\rho(x)}$ for the full derivative (valid over the unrestricted domain), this ambiguity can be expressed as

$$\left.\frac{\delta A[\rho]}{\delta\rho(x)}\right|_C = \frac{\delta A[\rho]}{\delta\rho(x)} + \lambda \frac{\delta C[\rho]}{\delta\rho(x)} \,, \tag{27}$$

provided, of course, that the full derivative exists. As has been proved in the Appendix of [46], in the chain rule of differentiation of a composite functional $A[\rho[q]]$, the full derivative $\frac{\delta A[\rho]}{\delta\rho(x)}$ may be replaced by any choice of the restricted derivative $\left.\frac{\delta A[\rho]}{\delta\rho(x)}\right|_C$,

$$\frac{\delta A[\rho[q]]}{\delta q(x)} = \int \left.\frac{\delta A[\rho]}{\delta\rho(x')}\right|_C \frac{\delta\rho(x')[q]}{\delta q(x)}\,dx' \,, \tag{28}$$

in the case $\rho(x)[q]$ is such that it satisfies the given constraint $C[\rho] = C$ for all $q(x)$'s – which is the case for Eqs.(21) and (22). It is crucial for both of the above cancellations of the ambiguity of restricted derivatives (yielding a unique $A[\rho+\delta_C\rho] - A[\rho]$ and a unique $\frac{\delta A[\rho[q]]}{\delta q(x)}$) that the integrals (in Eqs.(25) and (28)) are taken over the whole space. In the case of applications of a derivative $\frac{\delta A[\rho]}{\delta \rho(x)}$ where the ambiguity of the corresponding restricted derivatives under a given constraint does not cancel, the full derivative *cannot* be replaced by another choice $\left.\frac{\delta A[\rho]}{\delta \rho(x)}\right|_C$. We should keep in mind that only the full derivative is capable of delivering the correct change of $A[\rho]$ due to a change of its variable at a given point $x'$ induced by a change of a function $q(x)$ $\rho(x')$ depends on – either $\rho(x')[q]$ obeys some constraint or not. An additional term $+\lambda \frac{\delta C[\rho]}{\delta \rho(x)}$ just unnecessarily, and incorrectly, modifies the result given by $\frac{\delta A[\rho]}{\delta \rho(x)}$. To gain more insight into this, one may consider again the example of a composite function $g(x(t), y(t))$, with $(x(t), y(t))$ now obeying the constraint $x^2(t) + y^2(t) = c$, e.g. Under this constraint on $g$'s variables, the ambiguous restricted derivative $\left(\frac{\partial g(x,y)}{\partial x}, \frac{\partial g(x,y)}{\partial y}\right) + \lambda(2x, 2y)$ will correctly deliver, for any choice of $\lambda$, the full first-order change of $g$ due to a change in $(x, y)$ that is in accordance with the constraint – but *not* a partial first-order change, such as $\frac{\partial g}{\partial x}\frac{dx(t)}{dt}dt$.

Thus, we conclude that a correct local hardness measure may be delivered only by

$$\eta(\vec{r}) = \frac{\delta \mu[N[n], v[n]]}{\delta n(\vec{r})} f(\vec{r}) \ . \tag{29}$$

However, there is an irresolvable problem with evaluating $\frac{\delta \mu}{\delta n(\vec{r})}$, as will be pointed out in the following section. We should add here that, strictly, the above local quantity is not quite a local counterpart of *hardness*, since the fixation of $v(\vec{r})$ in $\mu[N, v]$ in obtaining $\eta[N, v]$ is an inherent part of the hardness concept. However, the full derivative that is not restricted by a fixed-$v(\vec{r})$ constraint *on the density domain* but still keeps $v(\vec{r})$ fixed is the derivative $\frac{\delta \mu[N[n], v]}{\delta n(\vec{r})}$, i.e. the derivative that yields Eq.(8). But due to its constancy, it is not capable of giving a local measure of hardness. It would only yield a local quantity that is proportional to the Fukui function itself, $\eta(\vec{r}) = \eta f(\vec{r})$; that is, it would actually measure regional *softnesses* by integration over molecular regions, considering Eq.(5). (We note that this is precisely the reason for the numerical observations of Torrent-Sucarrat et al. [18,19], who found that the regional integrals calculated with Eq.(8) used in Eq.(20) predict high regional hardness for actually soft regions – in the case of globally soft systems. This is then not surprising, since this is just what is expected from the Fukui function. The interesting fact, which gives the findings of Torrent-Sucarrat et al. high significance, is that this "local hardness" expression works well for hard systems, which implies that the Fukui function actually indicates local hardness instead of softness in the case of globally hard systems. Therefore, the interpretation of the Fukui function as a general local softness measure has to be reconsidered. But it is clear that $\eta(\vec{r}) = \eta f(\vec{r})$ also cannot be a local hardness measure.)

To close this section, it is worth exhibiting the ambiguity of the regional integrals Eqs.(20) and (23) that would be caused by the ambiguity of $\left.\frac{\delta\mu}{\delta n(\vec{r})}\right|_v$ and $\left.\frac{\delta\mu}{\delta n(\vec{r})}\right|_N$, respectively, if the use of those restricted derivatives, instead of the full derivative $\frac{\delta\mu}{\delta n(\vec{r})}$, was actually allowed in the case of integrals not covering the whole space. In the case of Eq.(23), the ambiguity of the restricted derivative appears in the form of a simple additive constant; that is, in the place of a given $\left.\frac{\delta\mu}{\delta n(\vec{r})}\right|_N$, any other $\left.\frac{\delta\mu}{\delta n(\vec{r})}\right|_N + \lambda$ can be taken as a choice for the chemical potential derivative over the $N$-restricted domain of $n(\vec{r})$'s. We may exhibit this ambiguity as

$$\left.\frac{\delta\mu}{\delta n(\vec{r})}\right|_N = \frac{\delta\mu}{\delta n(\vec{r})} + \lambda \ . \tag{30}$$

This ambiguity then leads to an ambiguity of $+\lambda\left(\frac{\delta N_\Omega}{\delta v(\vec{r})}\right)_N$ in $f_\Omega(\vec{r})$. The ambiguity Eq.(30) may be expressed with other particular choices of $\left.\frac{\delta\mu}{\delta n(\vec{r})}\right|_N$ replacing $\frac{\delta\mu}{\delta n(\vec{r})}$ in Eq.(30). Such a choice is

$$\left(\frac{\delta\mu[N,v[n]]}{\delta n(\vec{r})}\right)_N = \int\left(\frac{\delta\mu[N,v]}{\delta v(\vec{r}')}\right)_N \frac{\delta v(\vec{r}')}{\delta n(\vec{r})} d\vec{r}' = \int f(\vec{r}')\frac{\delta v(\vec{r}')}{\delta n(\vec{r})} d\vec{r}' \ , \tag{31}$$

which is the analogue of Eq.(8). With this, then, we may also write

$$\left.\frac{\delta\mu}{\delta n(\vec{r})}\right|_N = \int f(\vec{r}')\frac{\delta v(\vec{r}')}{\delta n(\vec{r})}d\vec{r}' + \lambda \qquad (32)$$

(emphasizing that $\lambda$ denotes an arbitrary constant throughout, not to be taken to be identical when appearing in different equations). By inserting Eq.(32) in Eq.(23), we obtain

$$f_\Omega(\vec{r}) = \int f(\vec{r}'')\int_\Omega \frac{\delta v(\vec{r}'')}{\delta n(\vec{r}')}\left(\frac{\delta n(\vec{r}')}{\delta v(\vec{r})}\right)_N d\vec{r}'d\vec{r}'' + \lambda\left(\frac{\delta N_\Omega}{\delta v(\vec{r})}\right)_N . \qquad (33)$$

(It can be seen that if $\Omega$ is chosen to be the whole space, Eq.(33) gives back the Fukui function.) As regards $\left.\dfrac{\delta\mu}{\delta n(\vec{r})}\right|_v$ in Eq.(20), it is determined only up to a term $+\int \lambda(\vec{r}')\dfrac{\delta v(\vec{r}')}{\delta n(\vec{r})}d\vec{r}'$ (with $\lambda(\vec{r})$ arbitrary), emerging from the fixed-$v(\vec{r})$ constraint, $v(\vec{r}')[n(\vec{r})] = v(\vec{r}')$ – which can be considered as an infinite number of constraints ("numbered" by $\vec{r}'$) on the $n(\vec{r})$-domain. This ambiguity may be exhibited as

$$\left.\frac{\delta\mu}{\delta n(\vec{r})}\right|_v = \frac{\delta\mu}{\delta n(\vec{r})} + \int \lambda(\vec{r}')\frac{\delta v(\vec{r}')}{\delta n(\vec{r})}d\vec{r}' , \qquad (34)$$

or with the particular choice Eq.(8) instead of $\dfrac{\delta\mu}{\delta n(\vec{r})}$, as

$$\left.\frac{\delta\mu}{\delta n(\vec{r})}\right|_v = \eta + \int \lambda(\vec{r}')\frac{\delta v(\vec{r}')}{\delta n(\vec{r})}d\vec{r}' . \qquad (35)$$

With Eq.(35), e.g., the ambiguity of Eq.(20) may then be given as

$$\eta_\Omega = \eta\int_\Omega f(\vec{r}')d\vec{r}' + \int \lambda(\vec{r}'')\int_\Omega \frac{\delta v(\vec{r}'')}{\delta n(\vec{r}')}f(\vec{r}')d\vec{r}'d\vec{r}'' . \qquad (36)$$

Eq.(35) gives back Eq.(17) with the choice $\lambda(\vec{r}) = f(\vec{r})$, as can be seen from Eq.(18). From Eq.(34), one can get back Eq.(15) if $-\lambda(\vec{r})$ is chosen to be a function $u(\vec{r})$ that integrates to 1, utilizing $\dfrac{\delta\mu}{\delta n(\vec{r})} = \int u(\vec{r}')\dfrac{\delta\mu}{\delta n(\vec{r})}d\vec{r}'$ and Eq.(9). This then shows that the possible choices of Eq.(14) are even more numerous than has been expected on the basis of Eq.(15).

**IV. Indeterminacy of the chemical potential's derivative with respect to the density**

For any possible application of Eq.(29), a proper method to evaluate the derivative of $v(\vec{r})$ with respect to the density is necessary, as revealed by Eq.(18). $v(\vec{r})$ is given as a functional of $n(\vec{r})$ by Eq.(9) itself; namely,

$$v(\vec{r}')[n] = \mu[n] - \frac{\delta F[n]}{\delta n(\vec{r}')} \ . \tag{37}$$

That is, in order to obtain the derivative of Eq.(37) with respect to $n(\vec{r})$ to determine $\frac{\delta \mu}{\delta n(\vec{r})}$ through Eq.(18), we already need to have $\frac{\delta \mu}{\delta n(\vec{r})}$! We cannot determine $\frac{\delta \mu}{\delta n(\vec{r})}$ without further information on $\mu[n]$, since from Eqs.(18) and (37),

$$\frac{\delta \mu[n]}{\delta n(\vec{r})} = \eta + \int f(\vec{r}') \frac{\delta}{\delta n(\vec{r})} \left( \mu[n] - \frac{\delta F[n]}{\delta n(\vec{r}')} \right) d\vec{r}' = \eta + \frac{\delta \mu[n]}{\delta n(\vec{r})} - \int f(\vec{r}') \frac{\delta^2 F[n]}{\delta n(\vec{r}) \delta n(\vec{r}')} d\vec{r}' \ , \tag{38}$$

which is an identity, involving Eq.(12).

$\mu$ is determined as a functional of the density by a boundary condition in Eq.(37). In the case of real, Coulombic, potentials, this will be according to the asymptotic condition $v(\infty) = 0$ on the external potentials, yielding

$$\mu[n] = \frac{\delta F[n]}{\delta n(\infty)} \ . \tag{39}$$

(Note that $n(\vec{r} \to \infty)$ is taken along one given direction, just as $v(\infty)$ needs to be fixed only along one direction – which then allows the extension to a wider domain of external potentials.) We emphasize that there is no other way to determine $\mu$ as a functional of $n(\vec{r})$ than the above, since $\mu$ (either as the chemical potential, i.e. the derivative of $E[N,v]$ with respect to $N$, or as the Lagrange multiplier in Eq.(9)) emerges directly as $\mu[N,v]$, which leaves $\mu[N[n],v[n]]$ undetermined, as seen above. With Eq.(39), then, we obtain

$$\frac{\delta \mu[n]}{\delta n(\vec{r})} = \frac{\delta^2 F[n]}{\delta n(\vec{r}) \delta n(\infty)} \ . \tag{40}$$

It is worth observing that Eq.(40) corresponds to the choice $u(\vec{r}') = \delta(\vec{r}' - \infty)$ in Eq.(15).

Eq.(40) seems to offer an easy way to evaluate $\frac{\delta \mu}{\delta n(\vec{r})}$: Just take the hardness kernel, and consider its limit as (any) one of its variables approaches infinity. However, a problem immediately arises. With using approximations for $F[n]$ that construct $F[n]$ simply in a form $F[n] = \int g(n(\vec{r}), \nabla n(\vec{r}), \nabla^2 n(\vec{r}), ...) d\vec{r}$ (which is common in practical calculations), delta functions $\delta(\vec{r} - \infty)$ appear as multipliers of constant components of the right of Eq.(40), which cannot yield a useful local index. One may argue that this is only an issue of the quality of approximation for $F[n]$, since as has been pointed out by Tozer et al. [54], a proper density functional $F[n]$ (if continuously differentiable) should yield an exchange-correlation potential that has a non-vanishing asymptotic value, which then may give a well-behaved Eq.(40). The problem, however, is more fundamental than this.

Consider the (exact) one-electron version of the DFT Euler-Lagrange equation Eq.(9),

$$\frac{\delta T_W[n]}{\delta n(\vec{r})} + v(\vec{r}) = -I , \qquad (41)$$

where $T_W[n]$ is the Weizsäcker functional $T_W[n] = \frac{1}{8}\int\frac{|\nabla n(\vec{r})|^2}{n(\vec{r})}d\vec{r}$, exactly valid for one-particle densities, while $I$ denotes the ionization potential, which is just minus the ground-state energy for one-particle systems. It is important that $T_W[n]$ is not only an exact functional for one-particle densities, in which case its derivative may differ from the generally valid $\frac{\delta F[n]}{\delta n(\vec{r})}$ by a ($n(\vec{r})$-dependent) constant, but in the zero-temperature grand canonical ensemble extension of the energy for fractional electron numbers [55] (see [56] for the spin-polarized generalization), it is the exact $F$ functional for densities with $N \leq 1$ [57], implying

$$\frac{\delta T_W[n_1]}{\delta n(\vec{r})} = \frac{\delta F[n_1]}{\delta n(\vec{r})}\bigg|_{-} \qquad (42)$$

(with no difference by a constant), and

$$-I(N=1) = \mu_{-}(N=1) , \qquad (43)$$

where the minus sign in the subscripts denotes that a left-side derivative is taken [in the zero-temperature ensemble scheme, the two one-sided derivatives are different in general, implying the existence of derivative discontinuities [55,56]]. We then obtain for $n(\vec{r})$'s with $N \leq 1$:

$$\mu[n] = \frac{\delta T_W[n]}{\delta n(\infty)} . \qquad (44)$$

However, the derivative of Eq.(44) with respect to $n(\vec{r})$,

$$\frac{\delta^2 T_W[n]}{\delta n(\vec{r})\delta n(\infty)} = -\frac{1}{4}\left(\frac{(\nabla n(\infty))^2}{(n(\infty))^3} - \frac{\nabla^2 n(\infty)}{(n(\infty))^2}\right)\delta(\infty-\vec{r}) + \frac{1}{4}\frac{\nabla n(\infty)}{(n(\infty))^2}\nabla\delta(\infty-\vec{r}) - \frac{1}{4}\frac{1}{n(\infty)}\nabla^2\delta(\infty-\vec{r}) \qquad (45)$$

(where by the arguments $\infty$, the corresponding asymptotic limits are meant), is ill-defined for electronic densities. The exponential asymptotic decay $e^{-2\sqrt{2I}\,r}$ [58,59] of such densities leads to infinite values of the factors of the delta functions above. (Note though that even without this, the delta functions would not make Eq.(45) a useful quantity.) More concretely speaking, the derivative of $\mu[n]$ does not exist, which can be seen by considering the infinitesimal increment $\delta\mu = \int \frac{\delta\mu[n]}{\delta n(\vec{r})} \delta n(\vec{r}) d\vec{r}$ of $\mu$ in a case where the ionization potential of $n(\vec{r})$ increases, i.e. the decay of $\tilde{n}(\vec{r}) = n(\vec{r}) + \delta n(\vec{r})$ is faster than $n(\vec{r})$'s. In such case, as can be checked readily, Eq.(45) leads to an infinite $\delta\mu$, whereas it should be $\tilde{I} - I$. We note that there is no *v*-representability issue (in the usual sense) regarding the differentiation of Eq.(44), since any (well-behaved) $n(\vec{r})$ delivers a corresponding $v(\vec{r})$ through Eq.(41), even if in many cases $n(\vec{r})$ will be an *excited-state* density corresponding to the delivered $v(\vec{r})$. We may add here that the Weizsäcker-functional derivative is not only a one-particle example, but $\frac{\delta T_W[n]}{\delta n(\vec{r})}$, a component of $\frac{\delta F[n]}{\delta n(\vec{r})}$ in the general case, *in itself* gives $-I$ (which equals $\mu_-$ [55]) in the case of finite electron systems, which can be seen if one inserts the density decay $e^{-2\sqrt{2I}\,r}$ [58,59] in $\frac{\delta T_W[n]}{\delta n(\vec{r})}$,

$$\frac{\delta T_W[n]}{\delta n(\vec{r})} = \frac{1}{8}\left(\frac{\nabla n(\vec{r})}{n(\vec{r})}\right)^2 - \frac{1}{4}\frac{\nabla^2 n(\vec{r})}{n(\vec{r})} \xrightarrow[r\to\infty]{} -I \ . \qquad (46)$$

It is important to point out that the above finding is not only some peculiar feature of the ensemble extension [55] of the energy for fractional *N*'s. In the case of other (possibly continuously differentiable) extensions, the derivatives of $T_W[n]$ and $F[n]$ may differ only by a (density-dependent) constant [46] at a one-particle density $n_1(\vec{r})$ (since the two functionals are equal for any $n_1(\vec{r})$), which then implies that their second derivatives may differ only by some $c(\vec{r}) + c(\vec{r}')$, as can be seen by applying (i) this constant-difference rule of derivatives to $\frac{\delta F[n_1]}{\delta n(\vec{r})} = \frac{\delta T_W[n_1]}{\delta n(\vec{r})} + C[n_1]$ itself and (ii) the symmetry property of second derivatives in $\vec{r}$ and $\vec{r}'$. Then, to obtain $\frac{\delta\mu[n_1]}{\delta n(\vec{r})}$ corresponding to a given fractional-*N* generalization of $F[n]$, $c(\vec{r}) + c(\infty)$ needs to be added to Eq.(45), where the function *c* depends on the generalization. Thus, the problematic Eq.(45) will still remain as a component of $\delta\mu/\delta n(\vec{r})$. Furthermore, for an *N*-conserving density variation, $\delta_N n(\vec{r})$, $c(\vec{r}) + c(\vec{r}')$ will cancel in $\delta\mu$, due to $\int \delta_N n(\vec{r}) d\vec{r} = 0$; consequently, we will find the same $\delta\mu$ for an *I*-increasing $\delta_N n(\vec{r})$ as above.

A very recent finding by Hellgren and Gross (HG) [60] gives further support of our conclusion regarding the ill-definedness of $\delta\mu/\delta n(\vec{r})$. These authors have showed that the right-side second derivative of the exchange-correlation (xc) component of $F[n]$ of the ensemble generalization for fractional $N$'s [55] diverges (exponentially) as $r \to \infty$, by which they have also placed earlier findings regarding the asymptotic divergence of the xc kernel [61] onto sound theoretical grounds. This divergent behaviour has been pointed out to emerge from the integer discontinuity of the xc kernel [60]. Since the left- and the right-side derivative at a given $n(\vec{r})$ may also differ only by a constant (see Appendix of [62] for a proof), the difference between the left- and the right-side second derivative may only be some $g(\vec{r}) + g(\vec{r}')$, on similar grounds as above (note that the left-side derivative and the right-side derivative of a functional at a given $n(\vec{r})$ may be considered as the derivatives of two different, continuously differentiable functionals that intersect on a subset of $n(\vec{r})$'s of a given $N$). HG has found that $g(\vec{r})$ of $g(\vec{r}) + g(\vec{r}') := \left.\frac{\delta^2 E_{xc}[n]}{\delta n(\vec{r})\delta n(\vec{r}')}\right|_+ - \left.\frac{\delta^2 E_{xc}[n]}{\delta n(\vec{r})\delta n(\vec{r}')}\right|_-$, which is the so-called discontinuity of the xc kernel at integer electron numbers, diverges exponentially as $r \to \infty$. Now, $F[n]$ is decomposed as $F[n] = T_s[n] + E_{xcH}[n]$, with $T_s[n]$ being the non-interacting kinetic-energy density functional and $E_{xcH}[n]$ the sum of $E_{xc}[n]$ and the classical Coulomb repulsion, or Hartree, functional. Since the latter is continuously differentiable, $E_{xcH}[n]$'s discontinuity properties are the same as $E_{xc}[n]$'s. For one-particle densities, we have $T_s[n_1] = T_W[n_1]$ and $E_{xcH}[n_1] = 0$. We emphasize that though $E_{xcH}[n]$ vanishes for one-particle densities, its derivative may still be a non-zero (spatial) constant for $n_1(\vec{r})$'s, and its second derivative may still be some $c(\vec{r}) + c(\vec{r}')$. Only the left-side (first and second) derivative of $E_{xcH}[n]$ of the ensemble fractional-$N$-extension [55] will be zero, in accordance with Eq.(42). Accordingly, $\left.\frac{\delta^2 E_{xcH}[n_1]}{\delta n(\vec{r})\delta n(\vec{r}')}\right|_+ = g(\vec{r})[n_1] + g(\vec{r}')[n_1]$. For this, the HG result also applies, as the addition of an electron to a one-electron system must be accounted for by similar features as adding an electron to a many-electron system; that is, $g(\vec{r})[n_1]$ diverges exponentially. Since the divergent behavior of $\left.\frac{\delta^2 E_{xcH}[n_1]}{\delta n(\vec{r})\delta n(\vec{r}')}\right|_+$ is closely related with long-range correlation effects [60,61], it is unlikely to be cancelled by $\left.\frac{\delta^2 T_s[n_1]}{\delta n(\vec{r})\delta n(\vec{r}')}\right|_+$

$$\left(=\left.\frac{\delta^2 T_W[n_1]}{\delta n(\vec{r})\delta n(\vec{r}')}\right|_+ +\left.\frac{\delta^2 (T_s[n_1]-T_W[n_1])}{\delta n(\vec{r})\delta n(\vec{r}')}\right|_+\right); \text{ consequently, } \left.\frac{\delta^2 F[n_1]}{\delta n(\vec{r})\delta n(\vec{r}')}\right|_+ \text{ diverges asymptotically,}$$

too. This then immediately gives that $\left.\dfrac{\delta\mu[n_1]}{\delta n(\vec{r})}\right|_+ = \left.\dfrac{\delta^2 F[n_1]}{\delta n(\vec{r})\delta n(\infty)}\right|_+$ is ill-defined, being infinite at every $\vec{r}$ ! Of course, similarly, the latter can be concluded from the divergence of the right-side xc kernel for the general $\left.\dfrac{\delta\mu[n]}{\delta n(\vec{r})}\right|_+$, too.

Thus, the unrestricted derivative of $\mu$ with respect to the density is ill-defined – at least, as long as we insist that the zero of energy should be fixed according to $v(\infty)=0$ for Coulombic potentials. If we chose some other, even though non-physical, fixation such as $\int g(\vec{r})v(\vec{r})d\vec{r}=0$, e.g. (where $g(\vec{r})$ is some fixed function that integrates to one and tends fast to zero with $\vec{r}\to\infty$), we would obtain $\mu[n]=\int g(\vec{r})\dfrac{\delta F[n]}{\delta n(\vec{r})}d\vec{r}$ generally for any potentials, which, then, would yield a proper derivative – but not of the *real* chemical potential. We refer to [63] for further insight into this issue and for a discussion of the related issue of the ground-state energy as a functional solely of the density.

Since the appearance of a preliminary version of the present work as an arXiv preprint (arXiv:1107.4249v4), a related study has been published by Cuevas-Saavedra et al. [64]. These authors deal with the problem of how to calculate the unconstrained local hardness Eq.(17) and conclude from similar contradictions as those pointed out in [63] that this local hardness concept is infinitely ill-conditioned and deduce further that it diverges exponentially fast asymptotically. Our conclusions thus go further, as we have shown that Eq.(17) is ill-defined everywhere for electronic systems.

**V. Local hardness as a constrained derivative with respect to the density**

It has thus been found that one cannot in principle obtain a local hardness measure by $\eta(\vec{r})=\left.\dfrac{\delta\mu[N[n],v[n]]}{\delta n(\vec{r})}\right|_v f(\vec{r})$, since one of the two mathematically allowed forms, Eq.(29), cannot be evaluated, while the other one, $\eta(\vec{r})=\eta f(\vec{r})$, is simply a measure of local softness in the case of soft systems. However, one may raise the question: Cannot we use Eq.(14) directly as a local hardness measure, irrespective of it being able to deliver a proper regional hardness concept or not? That is, one would not be interested in getting hardness values corresponding to regions of molecules, but only in obtaining a pointwise measure, which, besides, should deliver the global hardness (via Eq.(7)) – but not regional ones. Although this is a questionable concept, it seems to be plausible to consider Eq.(14) a proper local hardness measure due to its intuitive interpretation as a measure of how the chemical potential changes if the number of electrons is increased locally (by an infinitesimal amount) in a given external potential setting. Therefore, we will examine this option, too.

So, we are interested in finding a fixation of the ambiguity of Eq.(14) that would properly characterize the chemical potential change due to a density change when the density domain is restricted to densities corresponding to the given $v(\vec{r})$. This requires a proper modification of the unconstrained "gradient" $\frac{\delta\mu}{\delta n(\vec{r})}$, which leads us to the concept of constrained derivatives [65]. (Note the difference of the names "restricted derivative" and "constrained derivative" [46], which is not a canonized terminology yet – but the names should be different for these two mathematically, and also manifestly, different entities.) To see how this concept works, consider the case of the simple $N$-conservation constraint, $\int n(\vec{r})d\vec{r} = N$; i.e., the domain of $n(\vec{r})$'s is restricted to those integrating to a given $N$. The functional derivative $\frac{\delta A[n]}{\delta n(\vec{r})}$ is obtained from the first-order differential Eq.(25) (which delivers the first-order change of $A[n]$ for any variation $\Delta n(\vec{r})$ of $n(\vec{r})$) by inserting $\Delta n(\vec{r}') = \delta(\vec{r}' - \vec{r})$. That is, we obtain the functional derivative (i.e. gradient) by weighting all (independent) directions in the functional domain equally. In a case the functional domain is restricted by some constraint $C[n] = C$, the allowed directions are restricted by Eq.(26); consequently, $\delta(\vec{r}' - \vec{r})$ cannot be inserted in Eq.(25). We need to find a modification of $\delta(\vec{r}' - \vec{r})$ that is in accordance with the constraint. For the $N$-conservation constraint, this is achieved by $\delta_N(\vec{r}' - \vec{r}) = \delta(\vec{r}' - \vec{r}) - u(\vec{r}')$, giving $\Delta_N n(\vec{r}') = \int (\delta(\vec{r}' - \vec{r}'') - u(\vec{r}'))\Delta n(\vec{r}'')d\vec{r}''$ [65], where $u(\vec{r})$ is an arbitrary function that integrates to one. Inserting this $\Delta_N n(\vec{r}')$ in Eq.(25) and taking $\Delta n(\vec{r}'') = \delta(\vec{r}'' - \vec{r})$ yields the proper modification of a derivative $\frac{\delta A[n]}{\delta n(\vec{r})}$:

$\frac{\delta A[n]}{\delta_N n(\vec{r})} = \frac{\delta A[n]}{\delta n(\vec{r})} - \int u(\vec{r}') \frac{\delta A[n]}{\delta n(\vec{r}')} d\vec{r}'$. The key for obtaining the constrained derivative for a given constraint $C[n] = C$, thus, is to find the $\Delta_C n(\vec{r}')$'s that obey the constraint, i.e. $C[n + \Delta_C n] - C[n] = 0$.

Now, consider the domain determined by the fixed-$v(\vec{r})$ constraint. This domain of $n(\vec{r})$'s will be a very thin domain – literally; it will be a single chain of densities $n(\vec{r})[N,v]$, with only $N$ changing (non-degeneracy is assumed, of course, which is a basic requirement when dealing with $n(\vec{r})[v]$). Consequently, there is not much choice in writing a proper $\Delta_v n(\vec{r}')$. The only possible form is

$$\Delta_v n(\vec{r}') = \frac{\partial n(\vec{r}')[N,v]}{\partial N} \Delta N \ . \tag{47}$$

Inserting this in Eq.(25),

$$D(A)[n, \Delta_v n] = \frac{\partial A[n[N,v]]}{\partial N} \Delta N \tag{48}$$

arises via an application of the chain rule of differentiation. By utilizing $\Delta N = \int \Delta n(\vec{r}')d\vec{r}'$ and taking $\Delta n(\vec{r}') = \delta(\vec{r}' - \vec{r})$, from Eq.(48) we then obtain

$$\frac{\delta A[n]}{\delta_v n(\vec{r})} = \frac{\partial A[n[N,v]]}{\partial N} \tag{49}$$

as the constrained derivative corresponding to the $v(\vec{r})$-conservation constraint. Interestingly, though not surprisingly (considering the very restrictive nature of the fixed-$v(\vec{r})$ constraint), there is no ambiguity at all in this expression – contrary to the $N$-conserving derivative, e.g., where the freedom in the choice of $u(\vec{r})$ represents an ambiguity.

Thus, we obtain that the only mathematically allowed derivative of $\mu$ with respect to the density under the fixed-$v(\vec{r})$ constraint is

$$\frac{\delta \mu}{\delta_v n(\vec{r})} = \frac{\partial \mu[N,v]}{\partial N} \tag{50}$$

(that is, the $v(\vec{r})$-constrained, or "$v(\vec{r})$-conserving", derivative of the chemical potential with respect to the density is simply its partial derivative with respect to $N$). Note that $\mu[N[n[N,v]], v[n[N,v]]] = \mu[N,v]$. It turns out, thus, that the severe ambiguity of Eq.(14), embodied in Eq.(35), can be narrowed down to the single choice of $\lambda(\vec{r}) = 0$ – which is the constant local hardness of Eq.(8). (In other words: while the definition of a functional derivative leads to an ambiguity, Eq.(35), under a fixed-$v(\vec{r})$ constraint, this ambiguity disappears if one wishes to use this derivative *in itself* as a physical quantity, i.e. not only in integral expressions such as Eqs.(25) and (28).) This leads us to the conclusion that the local hardness concept defined by Eq.(14) *necessarily* gives the constant local hardness of Eq.(8), hence is not a proper basis to define a local measure of hardness. We can sum up our findings so far as: Here, we have shown that Eq.(8) is the only mathematically allowed choice if we wish to obtain a local hardness measure *directly by* Eq.(14), while previously we have shown that if we want to have a local hardness measure by $\left.\frac{\delta \mu}{\delta n(\vec{r})}\right|_{v} f(\vec{r})$, in order to have proper regional hardnesses as well, *then* the only allowed choices are $\eta f(\vec{r})$ and Eq.(29) – but the former cannot be a (general) local hardness measure because of its proportionality to the Fukui function.

## VI. The parallel problem of defining a local chemical potential

Defining a local hardness via Eq.(14) naturally raises the idea of defining a local counterpart of the chemical potential itself in a similar fashion. By a local counterpart of $\mu$, we mean a local index that indicates the local distribution of $\mu$ *within* a given ground-state system, i.e. not some $\vec{r}$–dependent chemical potential concept, like that of [6], that yields $\mu$ as its special, ground-state, case. We may introduce the following local quantity:

$$\tilde{\mu}(\vec{r}) = \left.\frac{\delta E[N[n], v[n]]}{\delta n(\vec{r})}\right|_{v} , \qquad (51)$$

which parallels Eq.(14). Of course, we then have the same kind of ambiguity problem as in the case of Eq.(14).

Fixing $v(r)$ as one of the variables of $E[N,v]$ will not yield a $\tilde{\mu}(\vec{r})$ that is a useful local measure of the chemical potential, similarly to Eq.(8), since this $\tilde{\mu}(\vec{r})$ will be constant in space – the chemical potential itself:

$$\tilde{\mu}(\vec{r}) = \left(\frac{\partial E[N,v]}{\partial N}\right)_{v(\vec{r})} \frac{\delta N}{\delta n(\vec{r})} = \mu . \qquad (52)$$

Eq.(52) may be obtained in another way as well, since the ground-state energy as a functional of the ground-state density can be obtained via two routes:

$$E[n] \equiv E[N[n], v[n]] \equiv E_{v[n]}[n] . \qquad (53)$$

The first route is through $E[N,v]$, while the second is through the energy density functional Eq.(10) of DFT – in both cases, the functional dependence of $v(\vec{r})$ on $n(\vec{r})$ is inserted into the corresponding places. Then, specifically, $\left.\frac{\delta E[n]}{\delta n(\vec{r})}\right|_v$ may be $\left(\frac{\delta E_v[n]}{\delta n(\vec{r})}\right)_v$, which equals $\mu$ on the basis of Eq.(9), giving back Eq.(52). We note here that the idea of a local chemical potential concept has been raised previously by Chan and Handy [66], as a limiting case of their more general concept of shape chemical potentials; however, they automatically took the energy derivative with respect to the density as the constant $\left(\frac{\delta E_v[n]}{\delta n(\vec{r})}\right)_v$, ignoring other possibilities. The constant local chemical potential concept of Eq.(52) is of course not without use; it may be considered as an equalized $\vec{r}$-dependent chemical potential, defined by $\mu(\vec{r}) \doteq \frac{\delta F[n]}{\delta n(\vec{r})} + v(\vec{r})$ [6]. The latter $\mu(\vec{r})$, however, is not a local chemical potential in the sense that it would be the local counterpart of a global property ($\mu$), but it is rather a kind of intensive quantity, which becomes constant when reaching equilibrium (here, ground state). Similar can be said of the $\vec{r}$-dependent, generalized hardness concept defined by Eq.(13) for general densities.

A general property of a $\tilde{\mu}(\vec{r})$ defined through Eq.(51) is

$$\mu = \int \left.\frac{\delta E[n]}{\delta n(\vec{r})}\right|_{v(\vec{r})} \left(\frac{\partial n(\vec{r})[N,v]}{\partial N}\right)_{v(\vec{r})} d\vec{r} = \int \tilde{\mu}(\vec{r}) f(\vec{r}) d\vec{r} \ ; \qquad (54)$$

i.e., it gives the chemical potential after integration when multiplied by the Fukui function – analogously to Eq.(7). We emphasize again that in spite of the great extent of ambiguity in Eq.(51), all choices will indeed give $\mu$ in Eq.(54), due to the fact that the density in $\left(\frac{\partial n(\vec{r})}{\partial N}\right)_v$ is varied with the external potential fixed, and in cases like this, the ambiguity of the inner derivative of the composite functional cancels out [46].

An appealing choice of the restricted derivative in Eq.(51) may be the unrestricted derivative,

$$\tilde{\mu}(\vec{r}) = \frac{\delta E[N[n], v[n]]}{\delta n(\vec{r})} \ . \qquad (55)$$

This quantity gives to what extent the ground-state energy changes when the density is changed by an infinitesimal amount at a given point in space. There will be places $\vec{r}$ in a given molecule where the same amount of infinitesimal change of $n(\vec{r})$ (at the given $\vec{r}$) would imply a greater change of the energy, while at other places, it would imply a smaller change in $E$, going together with a higher and a lower local value of $\tilde{\mu}(\vec{r})$, respectively. The most sensitive site of a molecule towards receiving an additional amount of electron (density) will be the site with the lowest value of $\tilde{\mu}(\vec{r})$, implying the biggest decrease of the energy due to an increase of the density at $\vec{r}$ by an infinitesimal amount – but only if the external potential changes accordingly. Eq.(55) can be evaluated as

$$\tilde{\mu}(\vec{r}) = \frac{\partial E[N,v]}{\partial N} + \int \frac{\delta E[N,v]}{\delta v(\vec{r}')} \frac{\delta v(\vec{r}')}{\delta n(\vec{r})} d\vec{r}' = \mu + \int n(\vec{r}') \frac{\delta v(\vec{r}')}{\delta n(\vec{r})} d\vec{r}' , \quad (56a)$$

or alternatively,

$$\tilde{\mu}(\vec{r}) = \frac{\delta E_v[n]}{\delta n(\vec{r})} + \int \frac{\delta E_v[n]}{\delta v(\vec{r}')} \frac{\delta v(\vec{r}')}{\delta n(\vec{r})} d\vec{r}' = \mu + \int n(\vec{r}') \frac{\delta v(\vec{r}')}{\delta n(\vec{r})} d\vec{r}' , \quad (56b)$$

where Eqs.(9) and (10) have been utilized. Note that the second term of Eq.(56) integrates to zero when multiplied by $f(\vec{r})$, as $v(\vec{r})$ is independent of $N$.

Eq.(55) is not only an appealing choice for Eq.(51), but on the basis of the argument given in the case of the local hardness in Sec.III, it is one of the two mathematically allowed choices to obtain a local chemical potential concept. The emerging local chemical potential is

$$\mu(\vec{r}) = \frac{\delta E}{\delta n(\vec{r})} f(\vec{r}) , \quad (57)$$

which gives regional chemical potentials via

$$\mu_\Omega = \int_\Omega \mu(\vec{r}) d\vec{r} . \quad (58)$$

(Just as in the case of Eq.(29), applying other choices of $\tilde{\mu}(\vec{r})$ of Eq.(51) in Eq.(57) would lead to an improper modification of the regional chemical potential values.) Unfortunately, however, the evaluation of $\frac{\delta E}{\delta n(\vec{r})}$ meets the same principal problem as the evaluation of $\frac{\delta \mu}{\delta n(\vec{r})}$. Inserting Eq.(37) in Eq.(56) gives

$$\frac{\delta E[n]}{\delta n(\vec{r})} = \mu + N \frac{\delta \mu[n]}{\delta n(\vec{r})} - \int n(\vec{r}') \frac{\delta^2 F[n]}{\delta n(\vec{r}) \delta n(\vec{r}')} d\vec{r}' , \quad (59)$$

which shows that the evaluation of $\mu[n]$'s derivative is required in order to determine $E[n]$'s derivative.

It is interesting to observe that the last term of Eq.(59) is just the original local hardness expression of Berkovitz et al., Eq.(16), times $N$. Eq.(59) indicates that small (positive) values of Eq.(16) imply that the global value $\mu$ is less decreased by them at the given points in space. This throws more light upon the recent finding [67] that Eq.(16) is a local indicator of *sensitivity* towards perturbations, which goes against the essence of the concept of local hardness. (The latter is not surprising in the view of Secs.III and V – actually nothing supports it as a formula for local hardness.)

The other possible way to obtain a local chemical potential measure is

$$\mu(\vec{r}) = \left(\frac{\delta E[N,v]}{\delta n(\vec{r})}\right)_v f(\vec{r}) = \mu f(\vec{r}) \; , \tag{60}$$

similar to the case of local hardness. In that case, $\eta(\vec{r}) = \eta f(\vec{r})$ could not give a correct local hardness measure since the Fukui function $f(\vec{r})$ is actually not an indicator of hard sites, while here the question is as to whether $f(\vec{r})$ can be considered an indicator of local electronegativity or not (note that $\mu$ is negative, and minus the chemical potential is the electronegativity). A positive answer would imply e.g. that two soft systems interact through their highest-local-electronegativity sites. However, to judge the appropriateness of such a possible role of $f(\vec{r})$, it should first be clarified what to expect from a local electronegativity concept – a matter well-worth of future studies.

Also just as in the case of the local hardness, one may examine the question as to what choices of Eq.(51) are allowed if one wishes to use Eq.(51) *itself* as a local chemical potential measure, ignoring the possibility of obtaining regional chemical potentials via $\mu_\Omega = \int_\Omega \tilde{\mu}(\vec{r}) f(\vec{r}) d\vec{r}$. Similarly as in Sec.V, it can be shown that actually the only possible choice to fix Eq.(51)'s ambiguity is given by the unique constrained derivative of the energy (with respect to the density) corresponding to the fixed-$v(\vec{r})$ constraint, which turns out to be

$$\frac{\delta E}{\delta_v n(\vec{r})} = \frac{\partial E[N,v]}{\partial N} \; , \tag{61}$$

i.e. the constant $\tilde{\mu}(\vec{r})$ of Eq.(52). That is, Eq.(51) cannot be taken as the direct definition of a local chemical potential, as it will only give back the chemical potential itself, which cannot be a *local* measure of the distribution of itself within a given species. Of course, as emphasized earlier, it can still be a special, equalized, case of a generalized, $\vec{r}$-dependent, chemical potential concept [6] – but it won't give a local reactivity index, characterizing molecular sites within individual species. (Note that Eq.(61) is not a trivial result obtained by the explicit fixation of $v(\vec{r})$ of $E[N[n],v]$, i.e. by $\left(\frac{\delta E[N[n],v]}{\delta n(\vec{r})}\right)_v$, but it is the derivative of $E[N[n],v[n]]$ with respect to $n(\vec{r})$ under the constraint of fixed $v(\vec{r})$.)

Finally, in parallel with Sec.III, we may consider the external-potential derivative of the energy,

$$\left(\frac{\delta E}{\delta v(\vec{r})}\right)_N = \int \frac{\delta E}{\delta n(\vec{r}')} \left(\frac{\delta n(\vec{r}')}{\delta v(\vec{r})}\right)_N d\vec{r}' . \tag{62}$$

External potential based reactivity indices have proved to be useful and have been much investigated [68]. The regional contributions to Eq.(62) are

$$n_\Omega(\vec{r}) = \int_\Omega \frac{\delta E}{\delta n(\vec{r}')} \left(\frac{\delta n(\vec{r}')}{\delta v(\vec{r})}\right)_N d\vec{r}' , \tag{63}$$

where we have utilized the fact that $\left(\frac{\delta E}{\delta v(\vec{r})}\right)_N$ is just the density. Eq.(63) gives a density component that can be viewed as the contribution of the given region $\Omega$ to $n(\vec{r})$. Here, an interesting possible application of Eq.(63) may be worth mentioning. A natural decomposition of the density is the one in terms of the occupied Kohn-Sham orbitals,

$$n(\vec{r}) = \sum_{i=1}^{N} |\phi_i(\vec{r})|^2 . \tag{64}$$

One may then look for regions $\Omega_i$ ($i=1,\ldots,N$) of the given molecule that contribute $n_i(\vec{r}) = |\phi_i(\vec{r})|^2$ to $n(\vec{r})$. Of course, this may imply a highly ambiguous result; however, the number of possible divisions of the molecule into $\Omega_i$ can be significantly reduced by searching for $\Omega_i$'s around the intuitively expectable regions where the single $n_i(\vec{r})$'s are dominant. In this way, one might find a spatial division of a molecule into subshells. This is probably an idea that is too speculative to be taken seriously, not to mention its practical evaluation, but is naturally suggested by Eq.(63). To go even further, one might assume that by applying the regions $\Omega_i$ found in this way in Eq.(23), the corresponding $f_{\Omega_i}(\vec{r})$'s might emerge to be $f_{\Omega_i}(\vec{r}) = \partial n_i(\vec{r})/\partial N$.

## VII. Conclusions

The traditional approach to defining a local measure of chemical hardness, by the derivative of the chemical potential with respect to the density subject to the constraint of a fixed external potential, has been re-examined. Although several problematic aspects of this approach, most importantly its ambiguity, had been pointed out before, it was still widely taken as a necessary framework to define a local hardness index. The ambiguity aspect is a negative feature as one needs to find the proper choice among the many possibilities, but at the same time, it gives hope that other concrete choice(s) to fix the ambiguity than those having various deficiencies may be found to serve better as a local hardness measure. However, we have shown in this study that the traditional approach is actually not ambiguous. The only mathematically allowed local hardness definitions emerging via that approach are (i) the one that gives the hardness itself in every point of space, and (ii) the one where the external potential constraint is actually dropped. In the latter case, however, the emerging local quantity is not yet the local hardness, but it should be multiplied by the Fukui function to get that. The first option arises as the unique constrained derivative corresponding to the fixed external potential constraint. The constancy of this quantity, however, makes it a useless concept as a local reactivity indicator. Although the local hardness concept emerging from the unrestricted chemical potential derivative (option (ii)) may be intuitively appealing, unfortunately it has been found that this concept is ill-defined, due to the fact that the chemical potential as a functional solely of the density is given by the asymptotic value of the derivative of the electronic internal energy density functional. Similar problems have been pointed out in defining a local chemical potential, as a *local* reactivity indicator, by the derivative of the ground-state energy with respect to the electron density. Our conclusion is that making the electron number local in the definitions of hardness and chemical potential, by substituting it with the electron density, is not a feasible approach to obtain local counterparts of these global reactivity descriptors; therefore, an essentially new way of defining corresponding local descriptors is necessary to be found.

**Acknowledgments:** The author acknowledges grants from the Netherlands Fund for Scientific Research and the U.S. Department of Energy TMS program (Grant No. DE-SC0002139).

**References**

[1] R. G. Parr and W. Yang, *Density Functional Theory of Atoms and Molecules* (Oxford University Press, New York, 1989).

[2] H. Chermette, J. Comput. Chem. **20**, 129 (1999); P. Geerlings, F. De Proft and W. Langenaeker, Chem. Rev. **103**, 1793 (2003); P. W. Ayers, J. S. M. Anderson, L. J. Bartolotti, Int. J. Quantum Chem. **101**, 520 (2005); M. H. Cohen and A. Wasserman, J. Phys. Chem. A **111**, 2229 (2007); J. L. Gázquez, J. Mex. Chem. Soc. **52**, 3 (2008); P. Geerlings and F. De Proft, Phys. Chem. Chem. Phys. **10**, 3028 (2008); S. B. Liu, Acta Physico-Chimica Sinica **25**, 590 (2009); R. K. Roy and S. Saha, Annu. Rep. Prog.

Chem., Sect. C: Phys. Chem. **106**, 118 (2010).

[3] L. Pauling, J. Am. Chem. Soc. **54**, 3570 (1932).

[4] R. S. Mulliken, J. Chem. Phys. **2**, 782 (1934).

[5] R. P. Iczkowski and J. L. Margrave, J. Am. Chem. Soc. **83**, 3547 (1961).

[6] R. G. Parr, R. A. Donnelly, M. Levy, and W. E. Palke, J. Chem. Phys. **68**, 3801 (1978).

[7] R. G. Pearson, J. Am. Chem. Soc. **85**, 3533 (1963).

[8] R. G. Parr and R. G. Pearson, J. Am. Chem. Soc. **105**, 7512 (1983).

[9] K. D. Sen (ed.), *Chemical Hardness*, Structure and Bonding **80** (Springer-Verlag, Heidelberg, 1993).

[10] R. G. Pearson, *Chemical Hardness: Applications from Molecules to Solids* (Wiley-VCH, Oxford, 1997).

[11] R. T. Sanderson, Science **114**, 670 (1951).

[12] P. K. Chattaraj, H. Lee, and R. G. Parr, J. Am. Chem. Soc. **113**, 1855 (1991).

[13] J. L. Gázquez, J. Phys. Chem. A **101**, 4657 (1997).

[14] J. L. Gázquez, J. Phys. Chem. A **101**, 9464 (1997).

[15] P. W. Ayers, J. Chem. Phys. **122**, 141102 (2005).

[16] P. W. Ayers, R. G. Parr, and R. G. Pearson, J. Chem. Phys. **124**, 194107 (2006).

[17] P. W. Ayers, Faraday Discuss. **135**, 161 (2007).

[18] R. G. Pearson, J. Chem. Educ. **64**, 561 (1987).

[19] R. G. Parr and P. K. Chattaraj, J. Am. Chem. Soc. **113**, 1854 (1991).

[20] R. G. Pearson, J. Chem. Educ. **76**, 267 (1999).

[21] P. W. Ayers and R. G. Parr, J. Am. Chem. Soc. **122**, 2010 (2000).

[22] M. Torrent-Sucarrat, J. M. Luis, M. Duran, M. Solà, J. Am. Chem. Soc. **123**, 7951 (2001)

[23] P. K. Chattaraj, P. W. Ayers, and J. Melin, Phys. Chem. Chem. Phys. **9**, 3853 (2007).

[24] W. T. Yang and R. G. Parr, Proc. Natl. Acad. Sci. USA **82**, 6723 (1985).

[25] M. Berkowitz, S. K. Ghosh and R. G. Parr, J. Am. Chem. Soc. **107**, 6811 (1985).

[26] S. K. Ghosh, Chem. Phys. Lett. **172**, 77 (1990).

[27] M. K. Harbola, P. K. Chattaraj, and R. G. Parr, Isr. J. Chem. **31**, 395 (1991).

[28] J. L. Gázquez, Struct. Bond. **80**, 27 (1993).

[29] P. K. Chattaraj, D. R. Roy, P. Geerlings, and M. Torrent-Sucarrat, Theor. Chem. Acc. **118**, 923 (2007).

[30] M. Torrent-Sucarrat, P. Salvador, M. Solà, and P. Geerlings, J. Comput. Chem. **29**, 1064 (2008).

[31] P. W. Ayers and R. G. Parr, J. Chem. Phys. **128**, 184108 (2008).

[32] S. Saha and R. K. Roy, Phys. Chem. Chem. Phys. **10**, 5591 (2008).

[33] T. Gál, P. Geerlings, F. De Proft, and M. Torrent-Sucarrat, Phys. Chem. Chem. Phys. **13**, 15003 (2011)   [arXiv:1104.3485].

[34] M. Torrent-Sucarrat, F. De Proft, P. Geerlings, and P. W. Ayers, Chem. Eur. J. **14**, 8652 (2008).

[35] M. Torrent-Sucarrat, F. De Proft, P. W. Ayers, P. Geerlings, Phys. Chem. Chem. Phys. **12**, 1072 (2010).

[36] Y. Li and J. N. S. Evans, J. Am. Chem. Soc. 117 (29), 7756 (1995).

[37] P. K. Chattaraj, J. Phys. Chem. A **105**, 511 (2001).

[38] J. Melin, F. Aparicio, V. Subramanian, M. Galván, and P. K. Chattaraj, J. Phys. Chem. A **108**, 2487 (2004).

[39] J. S. M. Anderson, J. Melin, and P. W. Ayers, J. Chem. Theory and Comp. **3**, 358 (2007).

[40] R. G. Parr and W. T. Yang, J. Am. Chem. Soc. **106**, 4049 (1984).

[41] D. Datta, J. Phys. Chem. **90**, 4216 (1986).

[42] P. W. Ayers, Chem. Phys. Lett. **438**, 148 (2007).

[43] See also P. W. Ayers and R. G. Parr, J. Chem. Phys. **129**, 054111 (2008).

[44] D. C. Ghosh and N. Islam, Int. J. Quantum Chem. **111**, 1961 (2010).

[45] P. W. Ayers, Theor. Chem. Acc. **118**, 371 (2007).

[46] T. Gál, J. Math. Chem. **42**, 661 (2007)   [arXiv:math-ph/0603027].

[47] R. G. Parr and L. J. Bartolotti, J. Phys. Chem. **87**, 2810 (1983).

[48] M. Berkowitz and R. G. Parr, J. Chem. Phys. **88**, 2554 (1988).

[49] P. K. Chattaraj, A. Cedillo, and R. G. Parr, J. Chem. Phys. **103**, 7645 (1995).

[50] For recent examples, see e.g. P. Mignon, S. Loverix, J. Steyaert, and P. Geerlings, Nucl. Acid. Res. **33**, 1779 (2005); A. Olasz, P. Mignon, F. De Proft, T. Veszpremi, and P. Geerlings, Chem. Phys. Lett. **407**, 504 (2005); A. S. Ozen, F. De Proft, V. Aviyente, and P. Geerlings, J. Phys. Chem. A **110**, 5860 (2006); P. Mignon, P. Geerlings, and R. A. Schoonheydt, J. Phys. Chem. C **111**, 12376 (2007); S. Saha and R. K. Roy, J. Phys. Chem. B **111**, 9664 (2007); S. Saha and R. K. Roy, J. Phys. Chem. B **112**, 1884 (2008).

[51] J. Garza and J. Robles, Int. J. Quantum Chem. **49**, 159 (1994); T. K. Ghanty and S. K. Ghosh, J. Phys. Chem. **98**, 9197 (1994); P. K. Chattaraj, A. Cedillo, and R. G. Parr, J. Chem. Phys. **103**, 10621 (1995); P. Fuentealba, J. Chem. Phys. **103**, 6571 (1995); S. B. Liu, F. De Proft, and R. G. Parr, J. Phys. Chem. A **101**, 6991 (1997); M. Torrent-Sucarrat, M. Duran, and M. Solà, J. Phys. Chem. A **106**, 4632 (2002); M. Torrent-Sucarrat, and P. Geerlings, J. Chem.

Phys. **125**, 244101 (2006); A. Borgoo, M. Torrent-Sucarrat, F. De Proft, and P. Geerlings, J. Chem. Phys. **126**, 234104 (2007).

[52] W. Langenaeker, F. De Proft, and P. Geerlings, J. Phys. Chem. **99**, 6424 (1995).

[53] C. J. Jameson and A. D. Buckingham, J. Chem. Phys. **73**, 5684 (1980), and references therein.

[54] D. J. Tozer, N. C. Handy, W. H. Green, Chem. Phys. Lett. **273**, 183 (1997); D. J. Tozer, Phys. Rev. A **56**, 2726 (1997); D. J. Tozer, N. C. Handy, J. Chem. Phys. **109**, 10180 (1998).

[55] J. P. Perdew, R. G. Parr, M. Levy, and J. L. Balduz, Phys. Rev. Lett. **49**, 1691 (1982).

[56] T. Gál and P. Geerlings, J. Chem. Phys. **133**, 144105 (2010)  [arXiv:0910.4782].

[57] E. Sagvolden and J. P. Perdew, Phys. Rev. A **77**, 012517 (2008).

[58] J. Katriel and E. R. Davidson, Proc. Natl. Acad. Sci. USA **77**, 4403 (1980).

[59] M. Levy, J. P. Perdew, and V. Sahni, Phys. Rev. A **30**, 2745 (1984).

[60] M. Hellgren and E. K. U. Gross, Phys. Rev. A **85**, 022514 (2012).

[61] O. Gritsenko and E. J. Baerends, J. Chem. Phys. **121**, 655 (2004).

[62] T. Gál, P. Ayers, F. D. Proft, and P. Geerlings, J. Chem. Phys. **131**, 154114 (2009).

[63] T. Gál, arXiv:1108.3865v2 (2011).

[64] R. Cuevas-Saavedra, N. Rabi, and P. W. Ayers, Phys. Chem. Chem. Phys. **13**, 19594 (2011).

[65] T. Gál, Phys. Rev. A **63**, 022506 (2001); J. Phys. A **35**, 5899 (2002); J. Phys. A **43**, 425208 (2010)  [arXiv:0708.1694].

[66] G. K.-L. Chan and N. C. Handy, J. Chem. Phys. **109**, 6287 (1998).

[67] P. W. Ayers, S. Liu, and T. Li, Phys. Chem. Chem. Phys. **13**, 4427 (2011).

[68] P. Fuentealba and R. G. Parr, J. Chem. Phys. **94**, 5559 (1991); P. Senet, J. Chem. Phys. **105**, 6471 (1996); R. Contreras, L. R. Domingo, J. Andrés, P. Pérez, and O. Tapia, J. Phys. Chem. A **103**, 1367 (1999); P. W. Ayers and M. Levy, Theor. Chem. Acc. **103**, 353 (2000); E. Chamorro, R. Contreras, and P. Fuentealba, J. Chem. Phys. **113**, 10861 (2000); P. W. Ayers and R. G. Parr, J. Am. Chem. Soc. **123**, 2007 (2001); C. Morell, A. Grand, and A. Toro-Labbé, J. Phys. Chem. A **109**, 205 (2005); C. Morell, A. Grand, and A. Toro-Labbé, Chem. Phys. Lett. **425**, 342 (2006); S. Liu, T. Li, and P. W. Ayers, J. Chem. Phys. **131**, 114106 (2009); N. Sablon, F. De Proft, P. W. Ayers, and P. Geerlings, J. Chem. Theory Comput. **6**, 3671 (2010); N. Sablon, F. De Proft, and P. Geerlings, Chem. Phys. Lett. **498**, 192 (2010).